# Model Potentials for a $C_{60}$ Shell

**S. T. Manson[1], A. S. Baltenkov[2], and A. Z. Msezane[3]**

[1]**Department of Physics and Astronomy, Georgia State University, Atlanta, Georgia 30303, USA**
[2]**Institute of Ion-Plasma and Laser Technologies, Tashkent 100125, Uzbekistan**
[3]**Center for Theoretical Studies of Physical Systems, Clark Atlanta University, Atlanta, Georgia 30314, USA**

**Abstract**: The spatial distribution of electric charges forming a square well potential has been analyzed. It is shown that this potential is created by two concentric spheres with a double layer of charges. A $C_{60}$ shell potential has been calculated under the assumption that it is formed by the averaged charge density of a neutral atom. It is further demonstrated that the phenomenological potentials simulating the $C_{60}$ shell potential belong to a family of potentials with a non-flat bottom. Two possible types of $C_{60}$ model potentials are proposed and their parameters have been calculated.

## 1. INTRODUCTION

Phenomenological approaches on the basis of simple model potentials for a $C_{60}$ shell are widely used to describe fullerene-like molecular systems, including endohedral fullerenes A@$C_{60}$, in which an atom A is encapsulated inside the hollow interior of the $C_{60}$ cage (see for example [1,2] and references therein). The experimental discovery of confinement resonances in the photoionization of the 4*d* subshell of Xe atom in molecular $\mathrm{Xe@C}_{60}^{+}$ [3] stimulated a number of theoretical investigations (see [4-10] and references therein). While some attempted to understand the predicted distortion of the 4*d* giant resonance [11] caused by the confinement effects, others attempted to describe the experimental results by varying the parameters $\Delta$ and $U_0$ of the phenomenological square well potential

$$U(r)=\begin{cases}-U_0, \text{if } R\text{-}\Delta/2\le r\le R+\Delta/2;\\ 0, \qquad \text{otherwise.}\end{cases} \tag{1}$$

Here, $\Delta = r_o - r_i$ is the thickness and $U_0$ is the depth of the square well potential; $R$ is the radius of the $C_{60}$ shell, i.e. the distance between the center of the fullerene cage and nuclei of the carbon atoms. The $C_{60}$ shell is formed by a superposition of the positive charge of atomic nuclei (or positive ions $C^{4+}$) and the negative charge of the electron cloud. It of interest to analyze the resultant spatial distribution of charge within the $C_{60}$ shell for the potential represented by Eq. (1) to inquire as to whether or not such a charge distribution is physically reasonable. In addition, it is also of interest to look at the potentials generated by other arrangements of the charge distribution of the atoms comprising the $C_{60}$ molecule.

Sections 2 and 3 of the paper are devoted to this analysis where two model potentials associated with the potential (1) are studied. In Sec. 4 the $C_{60}$ shell potential is discussed assuming that it is generated by the averaged (over the sphere of radius $R$) charge density of neutral atom. In Sec. 5 two possible types of the $C_{60}$ potential are considered and their parameters are calculated. In Sec. 6 the charge density forming a spherical potential well with zero thickness is analyzed. Section 7 gives conclusions.

## 2. SPATIAL DISTRIBUTION OF CHARGE

The energy of electron interaction with $C_{60}$ shell (1) is connected with the potential of electric field $\varphi(r)$ in which the electron is moving by the relation $U(r) = -\varphi(r)$. Here it is taken into account that the electron charge is equal to –1. The atomic units (au) ($\hbar = m = |e| = 1$) are used throughout this paper. The electrostatic field potential $\varphi(r)$ is defined by the Poisson equation

$$\Delta\varphi = -4\pi\rho\,, \qquad (2)$$

where $\rho(r)$ is the density of the electric charge forming the spherically symmetrical potential well (1). The radial dependence of this density is determined by

$$\frac{d^2}{dr^2}[rU(r)] = 4\pi r\rho(r)\,. \qquad (3)$$

Appling these general formulas to the square well potential (1), it is easy to see that the density of electric charge forming the potential well (1) is zero in all space excluding the epsilon neighborhoods of the radii $r_o$ and $r_i$ where the function $U(r)$ changes discontinuously from zero to $-U_0$ and *vise versa*. The charges forming the potential well (1) are smeared over the surfaces of two spheres with a gap Δ between them. More detailed information on the behavior of the function $\rho(r)$ near the surfaces of the concentric spheres of the radii $r_o$ and $r_i$ can be obtained if the function $U(r)$ is represented as two Fermi steps [4] which renders the potential continuous at the walls

$$U(r) = \begin{cases} -U_0[1+\exp\{(R-r-\Delta/2)/\eta\}]^{-1} \text{ for } r \le R, \\ -U_0[1+\exp\{(r-R-\Delta/2)/\eta\}]^{-1} \text{ for } r \ge R. \end{cases} \qquad (4)$$

Here, $\eta$ is the diffuseness parameter; potentials (1) and (4) coincide in the limit $\eta \to 0$. Substituting Eq. (4) into Eq. (3) we obtain for the charge density

$$4\pi r\rho(r) = U_0 \frac{1}{\eta^2} \times \begin{cases} \dfrac{z_+}{(1+z_+)^3}[r(1-z_+) - 2\eta(1+z_+)],\, r \le R; \\ \dfrac{z_-}{(1+z_-)^3}[r(1-z_-) - 2\eta(1+z_-)],\, r \ge R. \end{cases} \qquad (5)$$

Here $z_\pm = \exp[(\pm R \mp r - \Delta/2)/\eta]$. The calculated results for the functions $U(r)$ and $4\pi r\rho(r)$ are presented in Figs. 1 and 2. For the numerical calculations the following values of the potential well parameters were selected: $U_0 = 0.302$, $R$=6.84 and $\Delta = 1.9$ [4]. Fig. 2 demonstrates the evolution of the charge distribution with the diffuseness parameter varied within the range $0.05 \le \eta \le 0.2$. It is seen that for small $\eta$ approximately a half of the positive charges (carbon atom nuclei or positive ions $C^{4+}$) of the $C_{60}$ shell are located near the sphere with the radius $r_i$ while the rest of the positive charges are smeared over the sphere with the radius $r_o$. The positive charge of each of these spheres ($\rho > 0$) is compensated by two thin layers of negative charges ($\rho < 0$) formed by electrons. Thus, the spatial distribution of the charges forming the square well potential is a pair of two-layer charge sandwiches with a gap Δ between them.

The reason for such a behavior of the function $\rho(r)$ is that the potential $U(r)$ is formed *per se* by two Heaviside step functions. Their derivatives of the first and second order (see the Poisson equation) are the delta functions $\delta(r-r_i)$ and $\delta(r-r_o)$ and the derivatives of these delta functions $\delta'(r-r_i)$ and $\delta'(r-r_o)$, respectively.

It is noted that the $C_{60}$ shell is uncharged as a whole; therefore the total charge

$$Q = 4\pi \int_0^\infty \rho(r) r^2 dr = 0\,, \qquad (6)$$

where the charge density $\rho(r)$ is described by Eq. (5).

## 3. EXPONENTIAL-POWER POTENTIAL

Introduced in [12] and employed in Ref. [13], the exponential power potential

$$U(r) = -U_0 \exp[-(r-R)^p / w^p] \qquad (7)$$

allows the potential shape to be changed continuously from a Gaussian type ($p$=2) to a square-well type ($p \to \infty$). Note that the parameter $p$ in (7) is even. Using the potential (7) we can analyze how the potential well with a flat bottom is transformed into a potential with the cusp-shaped bottom. Substituting (7) into the Poisson equation (3), we obtain for the charge density

$$4\pi r\rho(r) = U_0 \frac{zyp}{w^p}[2 - ry\frac{p}{w^p} + r\frac{(p-1)}{(r-R)}]. \quad (8)$$

where the following replacements have been made,

$$z = \exp[-(r-R)^p / w^p]; \qquad y = (r-R)^{p-1}. \quad (9)$$

The calculated results for the functions $U(r)$ and $4\pi\rho(r)$ are presented in Fig. 3. For the numerical calculations the values of the potential well parameters ( $w = \Delta/2$ ) were the same as in Sec. 2. For $p$=2 we deal with the cusp-shaped bottom. In this potential well the atomic nuclei (or the positive ions) of the carbon atoms are localized near the sphere with the radius $R$, while the electronic clouds border the positively charged sphere inside and beyond it, i.e. the $C_{60}$ shell charges form a three-layer charge sandwich. The cusp-shaped bottom of the well becomes flatter with the increase of the $p$ parameter. The charge density distributions in Figs. 2 and 3 are qualitatively similar, providing evidence that the potential (1) can be created only if the assumption that the $C_{60}$ shell is formed by two double-charged spheres with the radii $r_i$ and $r_o$. Unfortunately, this does not appear to correspond to the real structure of the $C_{60}$ molecule.

## 4. AVERAGED DENSITY OF NEUTRAL ATOM CHARGE

A more realistic form of the potential well of the $C_{60}$ shell can be obtained as a result of averaging the charge density of a neutral atom over the sphere of radius $R$. For simplicity we will assume a one-electron atom. According to Ref. [14], the averaged density of atomic electrons is written as

$$\langle \rho_e(r) \rangle = -\frac{1}{4\pi}\int |\psi(\mathbf{r}-\mathbf{R})|^2 \, d\Omega. \quad (10)$$

At the center of the potential shell the electron density is finite and equal to

$$\langle \rho_e(r \to 0) \rangle = -|\psi(\mathbf{R})|^2. \quad (11)$$

Since the total charge of the electronic cloud is equal to unity we have the following formula

$$4\pi\int_0^\infty \langle \rho_e(r) \rangle r^2 dr = -1. \quad (12)$$

The potential created by this spherically symmetrical electronic cloud at the point **r** is

$$\varphi_e(r) = \int \frac{\langle \rho_e(r') \rangle}{|\mathbf{r}-\mathbf{r}'|} d\mathbf{r}' = 4\pi \left[ \frac{1}{r}\int_0^r \langle \rho_e(r') \rangle r'^2 dr' + \int_r^\infty \langle \rho_e(r') \rangle r' dr' \right]. \quad (13)$$

The electronic cloud potential in the center of the sphere is

$$\varphi_e(r \to 0) = 4\pi \int_0^\infty \langle \rho_e(r') \rangle r' dr'. \quad (14)$$

Far from the sphere center, namely for $r >> R$ we have

$$\varphi_e(r \to \infty) = 4\pi \frac{1}{r}\int_0^\infty \langle \rho_e(r') \rangle r'^2 dr' = -\frac{1}{r}. \quad (15)$$

Averaged over the sphere of radius $R$, the charge density of the atomic nucleus is

$$\langle \rho_n(r) \rangle = \frac{1}{4\pi}\int \delta(\mathbf{r}-\mathbf{R}) d\Omega = \frac{1}{4\pi}\frac{\delta(r-R)}{R^2}. \quad (16)$$

The potential created by this positive charge is obtained by substituting Eq. (16) into Eq. (13),

$$\varphi_n(r) = \begin{cases} 1/R & \text{for } r \le R, \\ 1/r & \text{for } r \ge R. \end{cases} \quad (17)$$

Thus, the potential of the electro-neutral layer formed by the neutral atom smeared over the sphere of the radius $R$

$$\varphi(r) = \varphi_n(r) + \varphi_e(r)\,, \tag{18}$$

which is finite at the center of the fullerene sphere and tends to zero for $r >> R$. The electronic wave function is written as

$$\psi(\mathbf{r}) = \frac{z^{3/2}}{\pi^{1/2}} \exp(-zr)\,. \tag{19}$$

The charge densities for different values of the parameter $z$ are presented in Fig. 4. The averaged density of the atomic nuclear charge (16) ( $\rho_n > 0$ ) is represented in Fig. 4 by the Lorentz curve

$$\langle \rho_n(r) \rangle = \frac{1}{4\pi R^2} \frac{1}{\pi} \frac{d}{(r-R)^2 + d^2}\,. \tag{20}$$

With increasing $z$ the electron cloud with $\rho_e < 0$ becomes more and more localized at both sides of the positively charged sphere with the radius $R$.

Calculated with the formulas (13) and (17), the potential wells $U(r) = -\varphi(r)$ formed by smeared atomic charge densities for different values of the parameter $z$ are presented in Fig. 5. All these potentials have the cusp-shaped bottoms. Hence, the phenomenological potentials modeling the $C_{60}$ shell potential are to be selected from a family of curves similar to those given in Fig. 5. It is evident that the number of such potentials is unlimited. We consider further two model potentials for the $C_{60}$ shell.

## 5. THE MODEL POTENTIALS

An important feature of the $C_{60}$ potential well is the presence of a shallow level in it. The electron affinity of $C_{60}$ determined by UV photoelectron spectroscopy [15] is about $I = 2.7 \pm 0.1$ eV. There are different data on the symmetry of the ground state of the $C_{60}^-$ ion. The symmetry group of the $C_{60}$ fullerene causes the degeneracy of an extra electron in this ion [16, 17]. According to the icosahedral ($I_h$) symmetry, the ground state of the $C_{60}^-$ ion is a *p*-like state. Also, experiment [18] shows that the electron attachment occurs in the ground state of *p*-symmetry. The fact that there is a threshold in the cross section for this process is explained by the existence of a centrifugal barrier that is absent for electron capture in an *s*-like state. In [19, 20], using a different method of measuring the slow electron attachment, the threshold of this process was not found. This establishes that the attachment of the slow electron is to the ground *s*-state. This observation is also consistent with the results obtained in [21] for the attachment of Rydberg electrons to $C_{60}$ demonstrating *s*-wave capture.

Both the different symmetries of the electron attachment were considered in the models of negative ion in Refs. [22, 23]. In the first case it was assumed that the ground state of the $C_{60}^-$ ion is a *p*-like state with the binding energy $E_p$= -2.7 eV. In the second one the ground state of the $C_{60}^-$ system was taken as an *s*-state with the binding energy $E_s$= -2.65 eV. The presence of these levels imposes limitations on the model potential parameters.

Consider the following model potentials. The Dirac-bubble potential $U(r) = -U_0 \delta(r-R)$ [22] can be considered as a limiting case of the Lorentz-bubble potential

$$U(r) = -U_0 \frac{1}{\pi} \frac{d}{(r-R)^2 + d^2} \tag{21}$$

for $d \to 0$. The maximum depth of the potential (21) at $r$=$R$ is $U_{\max} = U_0 / \pi d$. The thickness of the potential well $\Delta$ at the middle of the maximum depth is $\Delta = 2d$. With increasing $r$ the potential (21) decreases as $r^{-2}$. Along with (21) we consider the Cosh-bubble potential that decreases exponentially with $r$:

$$U(r) = -\frac{U_{\max}}{\cosh^n[\alpha(r-R)]}\,. \tag{22}$$

We further assume that $n$=1. In the middle of the maximum depth of the well (22), the thickness $\Delta$ is related to the parameter $\alpha$ as

$$\Delta = \frac{2}{\alpha}\ln(2+\sqrt{3}) = 2.633916/\alpha\,. \tag{23}$$

The parameters of these potential $U_{\max}$ and $\Delta$ should be connected with each other in such a way that in the potential wells (21) and (22) there is a *s*-like state or a *p*-like state with the specific energy $E=-I$. The parameters $U_{\max}$ and $\Delta$ are defined by solving the wave equation for the radial parts of the wave functions

$$\chi''_{nl} - \frac{l(l+1)}{r^2}\chi_{nl} + 2[E-U(r)]\chi_{nl} = 0\,, \tag{24}$$

where $\chi_{nl}(r) = rR_{nl}(r)$. For fixed thickness of the potential shell $\Delta$, the parameters $U_{\max}$ should provide the solutions $\chi_{nl}(r)$ of the wave equation (24), which decrease exponentially with *r*. A set of such pairs of parameters defines two families of potentials $U(r)$ where the *s*-like or *p*-like ground state with binding energy *E* exists. Eq. (24) with potentials (21) and (22) was solved numerically by the Runge-Kutta method [24]. The thickness $\Delta$ was set equal to 1, 2 and 3 au. The eigenvalue in Eq. (24) was set equal to *E*= -2.65 eV, when the radius *R*=6.665 [25]. The parameter $U_{\max}$ was varied until the eigenfunction $\chi_{nl}(r)$ vanished at large distances from the center of the $C_{60}$ cage. The calculated results for these functions for the orbital angular momentum *l*=0 (*s*-like ground state) and *l*=1 (*p*-like ground state) are presented in Fig. 6. For comparison, the wave functions calculated in [22, 23] with the Dirac-bubble potential ($\Delta=0$) are given in the same figures. With the increase of the parameter $\Delta$ the cusp-behavior of the wave function for zero-thickness changes to the smooth behavior of $R_{nl}(r) = \chi_{nl}(r)/r$ near the radius $r \approx R$. According to Fig. 6, the shapes of the wave functions depend comparatively weakly on the shape of the potential wells in which the electron is localized. The potentials with the calculated parameters $\Delta$ and $U_{\max}$ are shown in Fig. 7. As expected, the Cosh-bubble potential is localized in a narrower (as compared to the Lorentz type potential) region near the sphere of the radius *R*.

## 6. THE DIRAC-BUBBLE POTENTIAL

In Sec. 2 we discussed the connection of the charge density for potential (1) with the delta functions $\delta(r-r_i)$ and $\delta(r-r_o)$ as well as with the derivatives of these functions $\delta'(r-r_i)$ and $\delta'(r-r_o)$. We now analyze the charge density corresponding to the Lorentz-bubble potential (21). Substituting the potential (21) into the Poisson equation (3) we obtain the following expression for the density of charges forming the Lorentz-bubble potential

$$4\pi\rho(r) = U_0 \frac{2d}{\pi r[z^2+d^2]^3}[2z(z^2+d^2)+r(d^2-3z^2)]\,, \tag{25}$$

where $z = r - R$. The numerical calculations of the density (25) show that the charge density, as in Fig. 4, forms a three-layer sandwich: the middle layer is positively charged (atomic nuclei or positive ions $C^{4+}$) and the outer layers are negatively charged (electron clouds). In the limit $d \to 0$ all these spherical layers have zero thickness and radius equal to the skeleton radius *R,* i.e. the Dirac-bubble potential corresponds to the data on the molecular structure of the $C_{60}$ cage. The reason for such a behavior of the function $\rho(r)$ is that the charge density, in this case, is defined by the second derivative $\delta''(r-R)$ rather than the first derivative of the delta-functions as in Sec. 2.

## 7. CONCLUSIONS

It has been shown that the potential wells for the $C_{60}$ shell with a flat bottom is created by the spatial distribution of positive and negative charges localized on two concentric spheres, which is contrary to the real molecular structure of $C_{60}$. However, while the potentials with flat bottoms (particularly the square well potential) remain usable from a practical viewpoint as a fitting potential, users must keep in mind that the actual potential of $C_{60}$ has a cusp-shaped form. The two model potentials for the $C_{60}$ shell with a cusp-shaped bottom have been considered as well. The parameters of these potentials leading to the existence of bound states of the negative ion $C_{60}^{-}$ have been calculated.

However, we have a conundrum here. Experiments [26] seem to show that the $C_{60}$ potential really does have relatively compact definable walls. But the model potentials with walls at more or less well-defined radii (square-like types that we have explored in Sections 2 and 3) are produced by rather non-physical charge distributions. However, if one takes a reasonable distribution of the charge, based upon where the carbon nuclei are actually located (at the sphere with radius $R$), a cusp-shaped potential results. But the cusp-shaped potential cannot be considered as a potential with well-defined walls. In any case, this matter should be looked at further.

It should be noted, however that the representation of the fullerene shell potential as a model function $U(r)$ is essentially an idealization of the real potential of the $C_{60}$ cage and, therefore cannot be expected to describe (using this model function) all features of either the $C_{60}$ molecule or the endohedral atom $A@C_{60}$ correctly.

## ACKNOWLEDGMENTS

This work was supported by the Uzbek Foundation Award Ф2-ФА-Ф164. ASB is very grateful to Dr. I. Bitenskiy for useful comments. STM and AZM are supported by the U.S. DOE, Division of Chemical Sciences, Geosciences and Biosciences, Office of Basic Energy Sciences, Office of Energy Research.

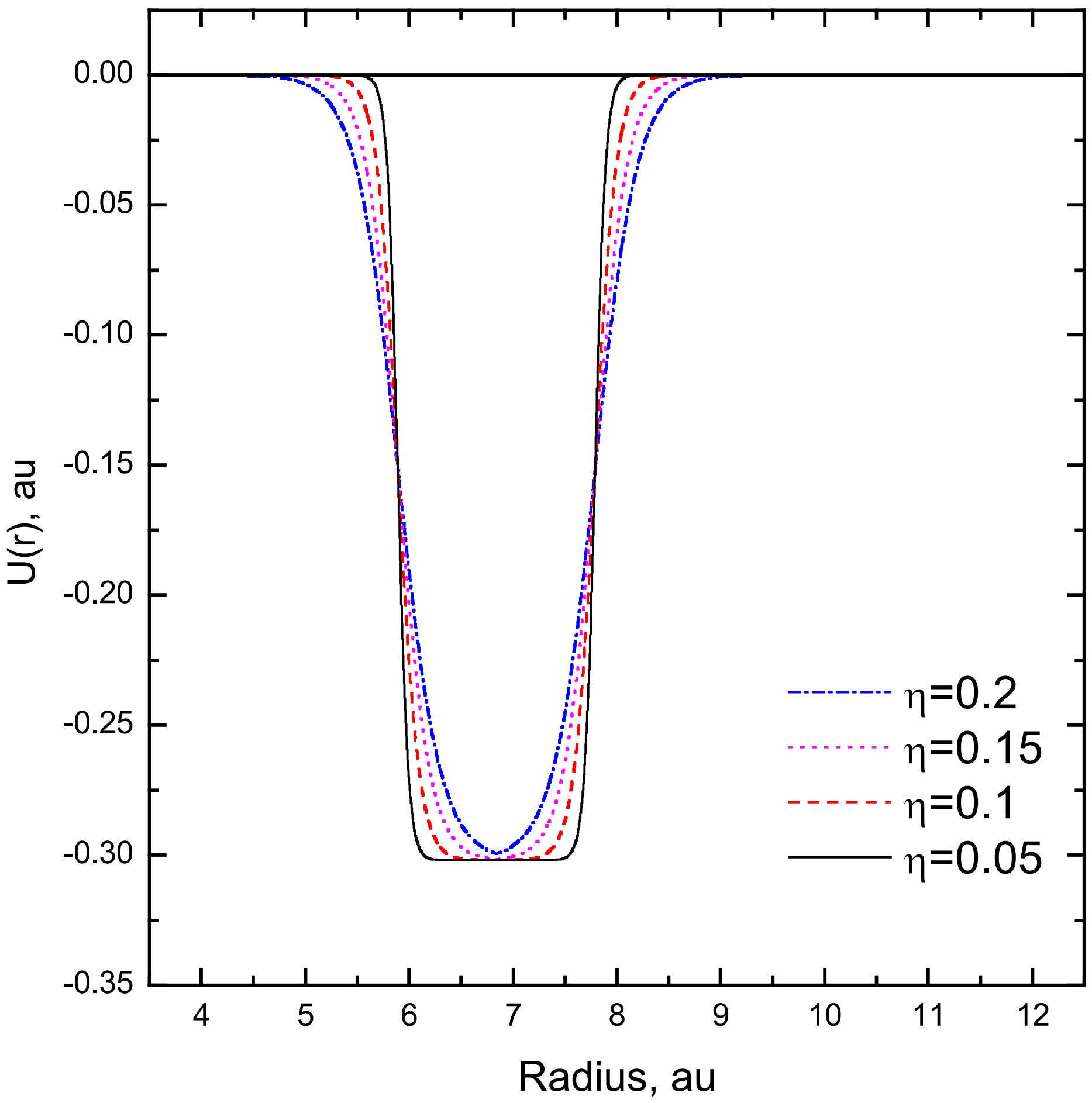

Fig. 1. The potential wells Eq. (4) for different diffuseness parameters

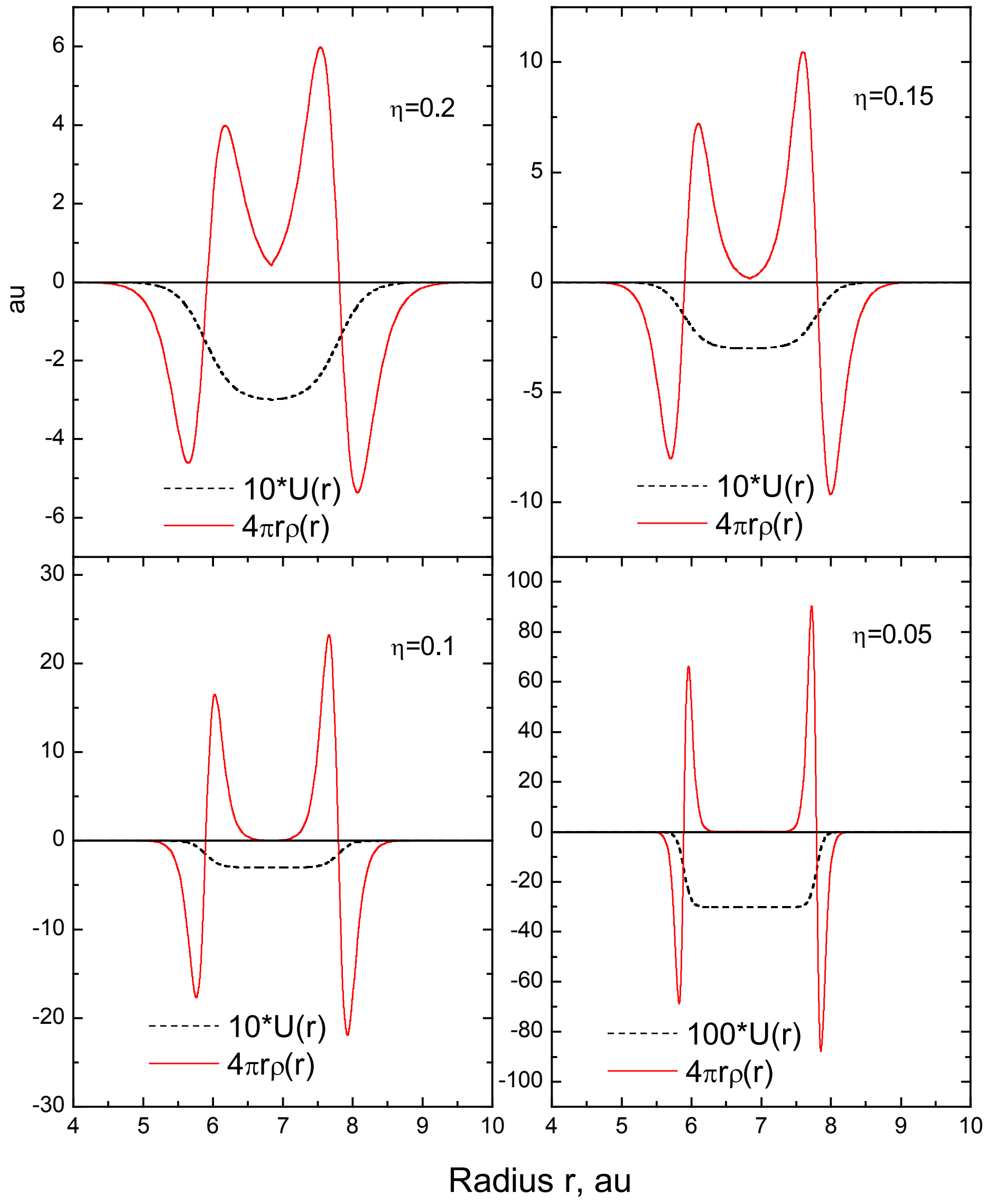

Fig. 2. The functions $U(r)$ and $4\pi r\rho(r)$ for different diffuseness parameters $\eta$ . Scaling factors for $U(r)$ are introduced for the both functions to be plotted in the same figure

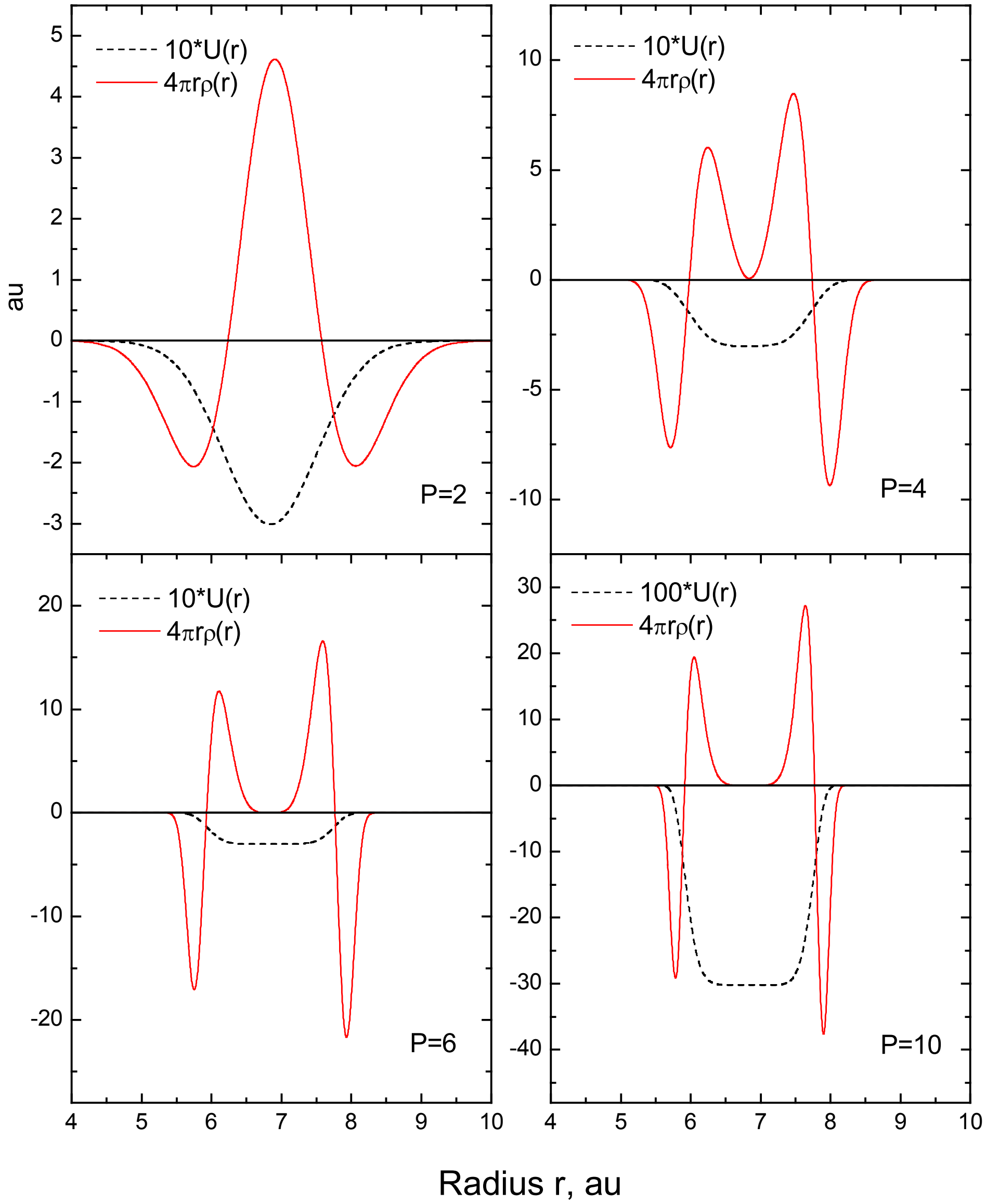

Fig. 3. The functions $U(r)$ (7) and $4\pi r\rho(r)$ (8) for different parameters $p$

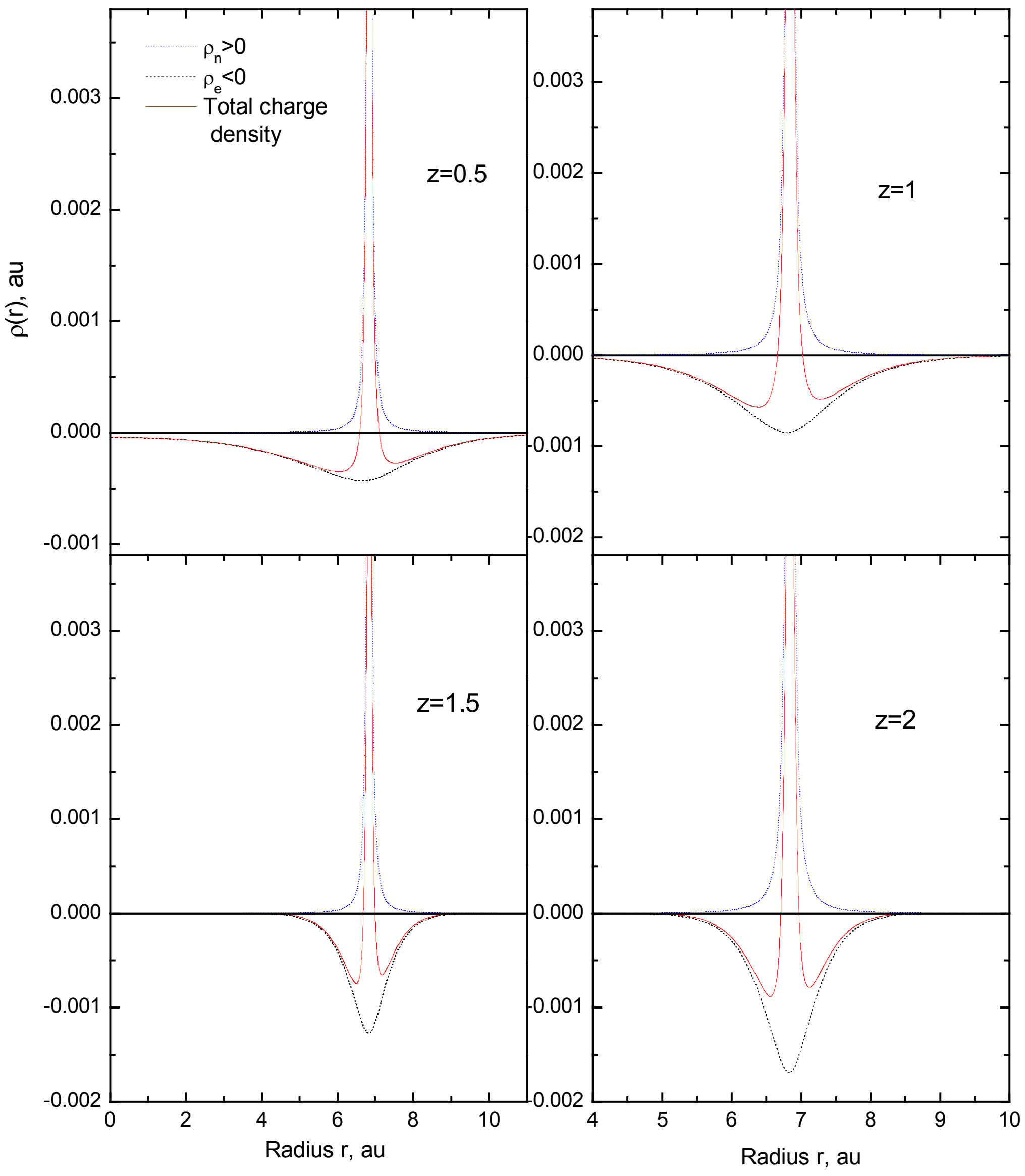

Fig. 4. The charge densities (10) and (20) for different parameters *z*

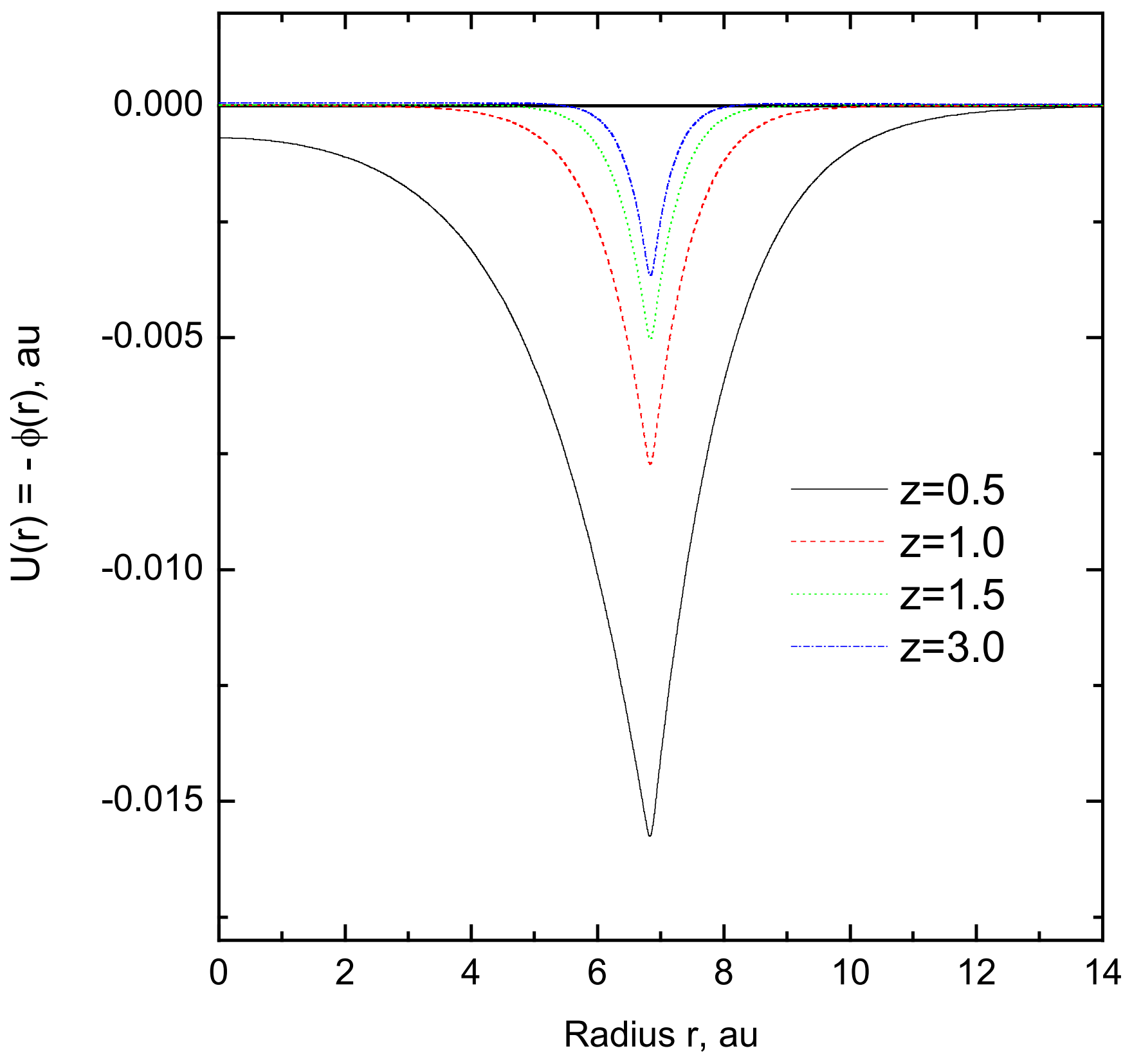

Fig. 5. The potential wells Eq. (18) with different *z*

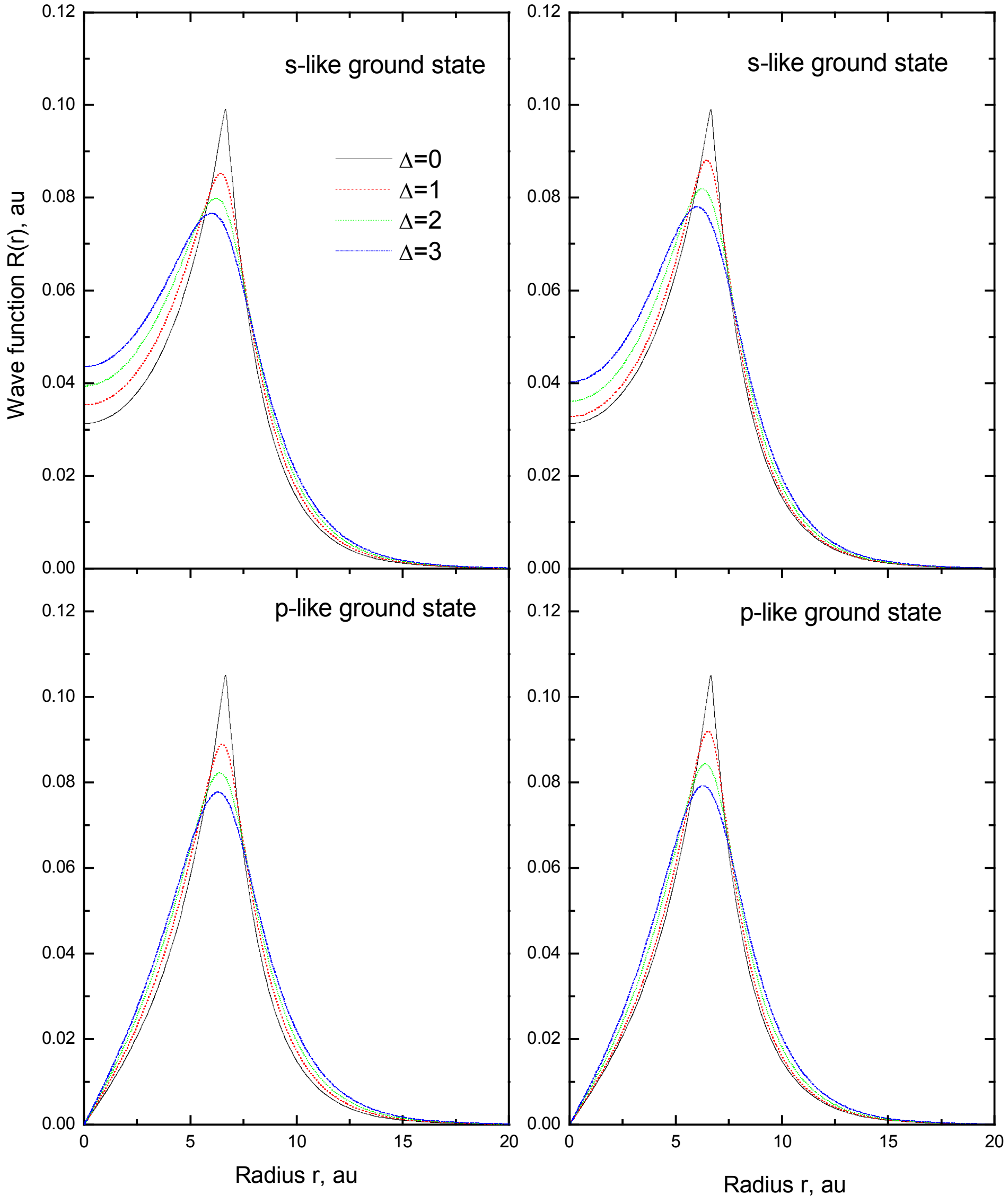

Fig. 6. The wave functions $R(r)$. The left panel is the Lorentz-bubble potential Eq. (21). The right panel is the Cosh-bubble potential Eq. (22)

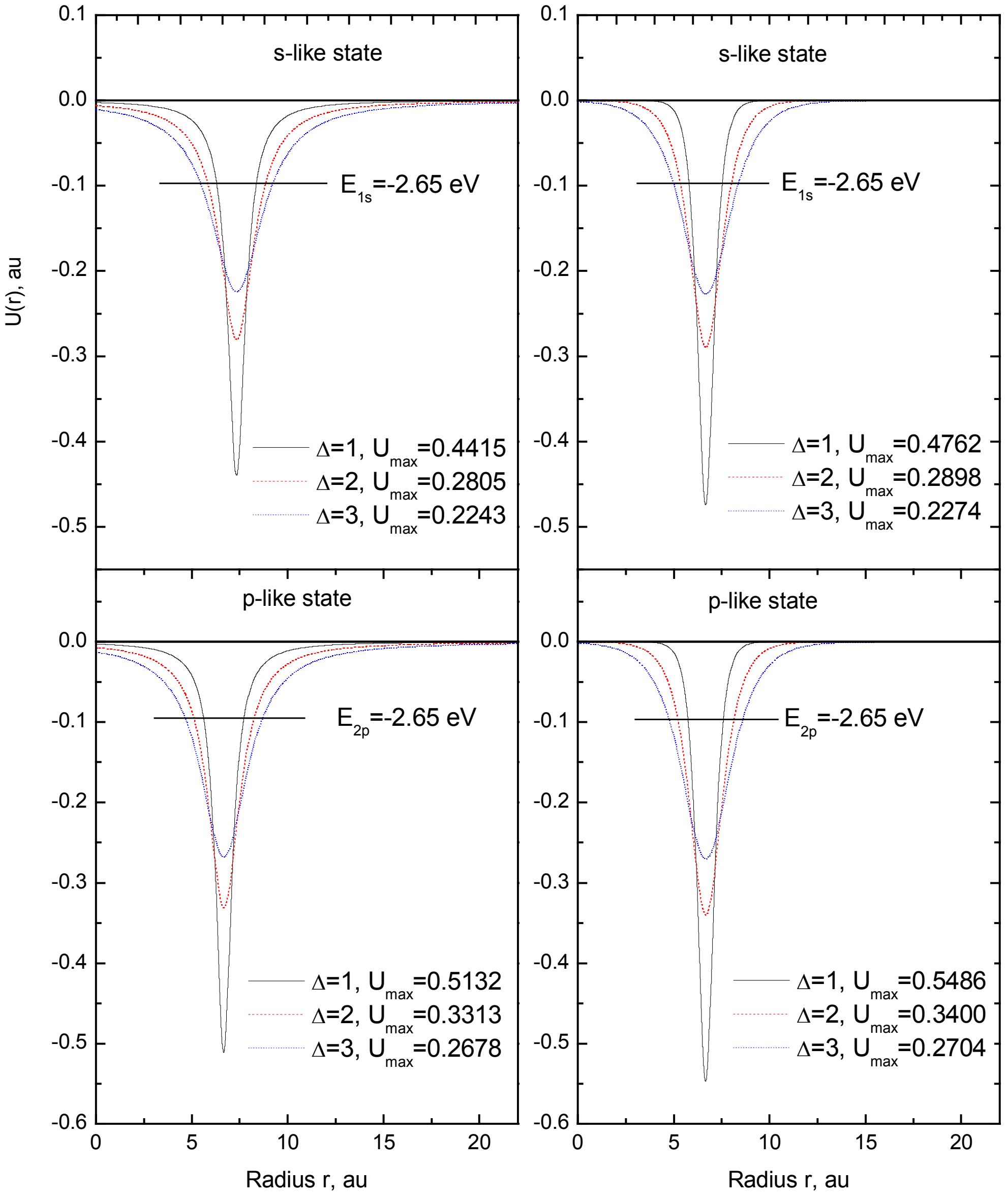

Fig. 7. The model potentials and their parameters. The left panel is the Lorentz-bubble potential Eq. (21). The right panel is the Cosh-bubble potential Eq. (22).